\documentclass{emulateapj}
\usepackage{graphics}
\usepackage{natbib}
\bibpunct{(}{)}{;}{a}{}{,}
\usepackage{amssymb}

\newcommand{\rh}{\varrho}
\newcommand{\divv}{\,\mbox{div}}
\newcommand{\grad}{\,\mbox{grad}}

\begin{document}

\title{Solar differential rotation and meridional flow:
  The role of a subadiabatic tachocline for the Taylor-Proudman balance}

\author{M. Rempel}

\affil{High Altitude Observatory,
       National Center for Atmospheric Research\footnote{The National
       Center for Atmospheric Research is sponsored by the National
       Science Foundation} , 
       P.O. Box 3000, Boulder, Colorado 80307, USA
      }

\email{rempel@hao.ucar.edu}

\shorttitle{Solar differential rotation and meridional flow}
\shortauthors{M. Rempel}

\begin{abstract}
We present a simple model for the solar differential rotation and meridional
circulation based on a mean field parameterization of the Reynolds stresses
that drive the differential rotation. We include the subadiabatic part
of the tachocline and show that this, in conjunction with turbulent
heat conductivity within the convection zone and overshoot region,
provides the key physics to break the Taylor-Proudman constraint, which 
dictates differential rotation with contour lines parallel to the 
axis of rotation in case of an isentropic stratification. 
We show that differential rotation with contour lines inclined by 
$10\degr - 30\degr$ with respect to the axis of rotation is a robust
result of the model, which does not depend on the details of the
Reynolds stress and the assumed viscosity, as long as the Reynolds stress 
transports angular momentum toward the equator. The meridional flow is more 
sensitive with respect to the details of the assumed Reynolds stress, but a 
flow cell, equatorward at the base of the convection zone and poleward 
in the upper half of the convection zone, is the preferred flow pattern.
\end{abstract}


\section{Introduction}
Helioseismology has revealed detailed information about the internal
rotation of the sun. Although the radiative interior of the Sun shows
a roughly uniform rotation, the convection zone shows a differential
rotation with a pole-equator difference of about $30\%$ of the
core rotation rate. The transition between the two regions is confined in a 
narrow shear layer, the so called tachocline, centered at about 
$0.7\,R_{\odot}$ and having a thickness of about $0.04\,R_{\odot}$
\citep{Charbonneau:etal:1999}. Within the convection zone the contours of 
constant angular velocity show in mid latitudes an inclination of about 
$25\degr$ with respect to the axis of rotation 
\citep{Schou:etal:1998,Schou:etal:2002}.

There is also robust observational evidence for a meridional flow, which
is poleward near the surface and has a flow velocity around 
$20\,\mbox{m}\,\mbox{s}^{-1}$.
This flow was first observed through movements of active regions and magnetic
filaments \citep{Labonte:Howard:1982,Topka:etal:1982} and later confirmed
through helioseismology \citep[see, e.g.,][]{Braun:Fan:1998,Haber:etal:2002,
Zhao:Kosovichev:2004}. Whereas most of the helioseismic studies focus only on
the flow field very close to the surface (using local helioseismology), 
\citet{Braun:Fan:1998} used global helioseismology to extend their analysis 
further down and found a poleward meridional flow in the entire upper 
half of the convection zone. Even though the return flow has not been detected 
yet, it is self-evident that it is located in the lower half of the convection 
zone because of mass conservation. Since the density is significantly larger 
at the base of the convection zone, the flow amplitude can be very small there,
on the order of a few $\mbox{m}\,\mbox{s}^{-1}$.
 
Following initial work by \citet{Glatzmaier:Gilman:1982} and 
\citet{Gilman:Miller:1986}, differential rotation has been addressed over the
past few decades by full spherical shell simulations of compressible 
convection. Even though recent results by \citet{Miesch:etal:2000}
and \citet{Brun:Toomre:2002} show improvement, the simulations still have
problems in reproducing the radial gradient of the differential within the 
convection zone. Although the models predict the magnitude of the differential
rotation (latitudinal variation) correctly, the contour lines of constant
angular velocity are still very close to the Taylor-Proudman state with
isolines parallel to the axis of rotation. The three-dimensional simulations 
also have 
problems in generating the observed meridional flow, which shows in the
upper half of the convection zone a poleward flow of an amplitude of about 
$20\,\mbox{m}\,\mbox{s}^{-1}$.  In contrast to observations, the 
three-dimensional simulations 
typically show several smaller flow cells of opposite sign and large temporal 
variability.   

A completely different approach is based on axisymmetric mean field models 
that parametrize the turbulent angular momentum transport
\citep[$\Lambda$-effect;][]{Kitchatinov:Ruediger:1993}. 
As discussed by \citet{Ruediger:etal:1998},
these models typically show solutions close to the Taylor-Proudman state
unless a very large value of the turbulent viscosity is used.

\citet{Kitchatinov:Ruediger:1995} showed  that an anisotropic convective
energy transport can produce a latitudinal variation in the temperature
that is large enough to overcome the Taylor-Proudman constraint and allow 
for solar-like differential rotation while assuming reasonable values
for the turbulent viscosity. The anisotropy of the energy flux results from
the rotational influence on the turbulence modifying the turbulent heat
diffusivities \citep{Kitchatinov:etal:1994}. Recently this model has been 
applied by \citet{Kueker:Stix:2001} to study the evolution of the solar 
differential rotation of the Sun from the pre-main-sequence Sun to
the present Sun by solving the momentum equations together with an 
mixing-length approach for the convective energy flux. 
The role of latitudinal variations of the entropy for the differential rotation
was also studied by \citet{Durney:1999,Durney:2003} within the framework of
mean field models. 

In this paper we present a model that is along the lines of the axisymmetric
mean field models mentioned above. Whereas the investigations of 
\citet{Kitchatinov:Ruediger:1995} and \citet{Kueker:Stix:2001} focused on
the role of an anisotropic convective energy flux, we focus our 
investigation on the role of a subadiabatic tachocline for the Taylor-Proudman
balance of the differential rotation. The aim of this paper is not to present
a complete model for the solar differential rotation, but rather a simple 
approach to investigate the specific question of 
how a subadiabatic tachocline, in conjunction with turbulent
heat conductivity within the convection zone and overshoot region,
can break the Taylor-Proudman constraint which requires a differential 
rotation constant on cylinders in case of an isentropic stratification.

\section{Model}
In this investigation we use a simplified model for studying differential
rotation and meridional circulation in the solar convection zone.
The basic assumptions underlying this approach are as follows.
\begin{enumerate}
  \item Axisymmetry and a spherically symmetric reference state.
  \item All processes on the convective scale are
    parameterized, leading to turbulent viscosity, turbulent heat conductivity,
    and turbulent angular momentum transport. 
  \item The equations can be linearized assuming $\rh_1\ll\rh_0$ and 
    $p_1\ll p_0$, and the reference state is assumed to be spherically 
    symmetric. Here $\rh_0$
    and $p_0$ denote the reference state values, whereas $\rh_1$ and $p_1$
    are the perturbations caused by the presence of differential rotation.
    The equations we solve are the fully  compressible, linearized,
    axisymmetric hydrodynamic equations.
  \item  The entropy equation includes only perturbations associated with 
    differential rotation. Therefore, the reference state is assumed to be
    in an energy flux balance. It is further assumed that the effect of
    convection (in the convection zone and the overshoot region) on entropy 
    perturbations associated with the differential rotation is purely 
    diffusive.
  \item The tachocline at the base of the solar convection zone is forced
    by a uniform rotation boundary condition at $r=0.65\,R_{\odot}$. 
\end{enumerate}
We emphasize that this model is not intended to be a complete
solar convection zone model, since fundamental processes required for
differential rotation such as turbulent angular momentum transport are 
parameterized. 

This model is also intended to be a basis for a 'dynamic' flux-transport 
dynamo including the {\boldmath$j\times B$} force 
feedback on meridional flow and differential rotation (future work).  

\subsection{Basic equations}
For this model we use the axisymmetric, fully compressible hydrodynamic 
equations. Since the perturbation of pressure and density caused by the 
differential rotation are small compared with the reference state values
[$\rh_1/\rh_0\sim p_1/p_0\sim (\Delta\Omega R_{\odot}/c_s)^2\sim 10^{-5}$],
we linearize the equations, assuming $\rh_1\ll\rh_0$ and $p_1\ll p_0$.
Since we do not use the anelastic approximation here (see section
\ref{numerics} for more details), we keep the time derivative in the
continuity equation:  
\begin{eqnarray}
   \frac{\partial \rh_1}{\partial t} &=& -\frac{1}{r^2}
      \frac{\partial}{\partial r}\left(r^2 v_r\rh_0\right)
      -\frac{1}{r\sin\theta}\frac{\partial}{\partial \theta}
      \left(\sin\theta v_{\theta}\rh_0\right)\;,\label{dens}\\
   \frac{\partial v_r}{\partial t} &=& -v_r\frac{\partial v_r}{\partial r}
      -\frac{v_{\theta}}{r}\frac{\partial v_r}{\partial \theta}
      +\frac{v_{\theta}^2}{r}-\frac{1}{\rh_0}\left(\rh_1 g(r) 
      +\frac{\partial p_1}{\partial r}\right)\nonumber \\
      && +\left(2\Omega_0\Omega_1+\Omega_1^2\right)r\sin^2\theta 
      +\frac{F_r}{\rh_0} \;,\label{vrad}\\
   \frac{\partial v_{\theta}}{\partial t} &=& 
      -v_r\frac{\partial v_{\theta}}{\partial r}
      -\frac{v_{\theta}}{r}\frac{\partial v_{\theta}}{\partial \theta}
      -\frac{v_r v_{\theta}}{r}
      -\frac{1}{\rh_0}\frac{1}{r}\frac{\partial p_1}{\partial\theta}
      \nonumber \\
      && +\left(2\Omega_0\Omega_1+\Omega_1^2\right)r\sin\theta\cos\theta 
      +\frac{F_{\theta}}{\rh_0} \;,\label{vthe} \\
   \frac{\partial \Omega_1}{\partial t} &=& 
      -\frac{v_r}{r^2}\frac{\partial}{\partial r}\left[r^2(\Omega_0+\Omega_1)
      \right]\nonumber\\
      &&-\frac{v_{\theta}}{r \sin^2\theta}\frac{\partial}{\partial \theta}
      \left[\sin^2\theta(\Omega_0+\Omega_1)\right]
      +\frac{F_{\phi}}{\rh_0 r\sin\theta} \;,\label{omeg}\\
   \frac{\partial s_1}{\partial t} &=& -v_r\frac{\partial s_1}{\partial r}
      -\frac{v_{\theta}}{r}\frac{\partial s_1}{\partial \theta}
      +v_r\frac{\gamma\delta}{H_p}+\frac{\gamma-1}{p_0}\mbox{Q}
      \nonumber\\
      &&+\frac{1}{\rh_0 T_0}\divv(\kappa_t\rh_0 T_0\grad s_1)\;,
      \label{entr}
\end{eqnarray}
where 
\begin{eqnarray}
  p_1&=&p_0\left(\gamma\frac{\rh_1}{\rh_0}+s_1\right)\;,\label{pressure}\\
  H_p&=&\frac{p_0}{\rh_0 g}\;.
\end{eqnarray}
We use here the dimensionless entropy $s=\ln(p/\rh^{\gamma})$, meaning that the
entropy equation Eq. (\ref{entr}) was made dimensionless by division through
$c_v=(\gamma-1)^{-1}{R}/\mu$.

The third term on the right-hand side of Eq. (\ref{entr}) describes
the effects of a non-adiabatic reference state, where 
$\delta=\nabla-\nabla_{\rm ad}$ is related to the gradient of the
reference state entropy through:
\begin{equation}
  \frac{ds_0}{dr}=-\frac{\gamma \delta}{H_p}\;.
\end{equation}
The fourth term in Eq. (\ref{entr}) considers the energy transfer through 
Reynolds stress that will be defined in detail in the following paragraph.
The last term describes turbulent diffusion of entropy perturbation
within the convection zone, where $\kappa_t$ denotes the turbulent
thermal diffusivity. We neglect in Eq. (\ref{entr}) the contribution
of the radiative energy flux, since even for overshoot values the turbulent
heat conductivity exceeds the radiative one by several orders of magnitude.

The viscous force {\boldmath$F$} follows from
\begin{eqnarray}
  {F}_r&=&\frac{1}{r^2}\frac{\partial}{\partial r}\left(r^2
     {R}_{rr}\right)+\frac{1}{r\sin\theta}\frac{\partial}
     {\partial \theta}\left(\sin\theta{R}_{\theta r}\right)
     \nonumber\\&&
     -\frac{{R}_{\theta\theta}+{R}_{\phi\phi}}{r} \;,\\
  {F}_{\theta}&=&\frac{1}{r^2}\frac{\partial}{\partial r}\left(r^2
     {R}_{r\theta}\right)+\frac{1}{r\sin\theta}\frac{\partial}
	 {\partial\theta}\left(\sin\theta{R}_{\theta\theta}\right)
	 \nonumber\\&&
	 +\frac{{R}_{r\theta}-{R}_{\phi\phi}\cot\theta}{r}\;,\\
  {F}_{\phi}&=&\frac{1}{r^2}\frac{\partial}{\partial r}\left(r^2
     {R}_{r\phi}\right)+\frac{1}{r\sin\theta}\frac{\partial}
     {\partial\theta}\left(\sin\theta{R}_{\theta\phi}\right)
     \nonumber\\&&
     +\frac{{R}_{r\phi}+{R}_{\theta\phi}\cot\theta}{r} \;,    
\end{eqnarray}
with the Reynolds stress tensor
\begin{equation}
  {R}_{ik}=-\rh_0<v_i^{\prime} v_k^{\prime}>=\nu_t\rh_0 
    \left({E}_{ik}-\frac{2}{3}\delta_{ik}\divv\,\mbox{\boldmath$v$}
        + \Lambda_{ik}\right)\;.\label{stresstensor}
\end{equation} 
Here ${E}_{ik}=v_{i;k}+v_{k;i}$ denotes the deformation tensor, which is
given in spherical coordinates by
\begin{eqnarray}
  {E}_{rr}&=&2\frac{\partial v_r}{\partial r}\;,\\
  {E}_{\theta\theta}&=&2\frac{1}{r}\frac{\partial 
    v_{\theta}}{\partial \theta}+2\frac{v_r}{r}\;,\\
  {E}_{\phi\phi}&=&\frac{2}{r}\left(v_r+v_{\theta}\cot\theta\right)\;,\\
  {E}_{r\theta}&=& {E}_{\theta r}=
    r\frac{\partial}{\partial r}\frac{v_{\theta}}{r}
    +\frac{1}{r}\frac{\partial v_r}{\partial \theta}\;,\\
  {E}_{r\phi}&=&{E}_{\phi r}=
    r\sin\theta\frac{\partial\Omega_1}{\partial r} \;,\\
  {E}_{\theta\phi}&=&{E}_{\phi\theta}= 
    \sin\theta\frac{\partial\Omega_1}{\partial \theta}\;,  
\end{eqnarray}
whereas $\Lambda_{ik}$ denotes the non diffusive Reynolds stresses, which are
responsible for driving differential rotation. We will discuss these terms 
later.

The amount of energy that is converted by the Reynolds stress is given
by:
\begin{equation}
  {Q}=\sum_{i,k}\frac{1}{2}{E}_{ik}{R}_{ik}\;.
\end{equation}
${Q}$ contains a heating term resulting from the dissipation
of kinetic energy through the dissipative contribution to the Reynolds stress
(terms proportional to ${E}_{ik}$) and a cooling term resulting from
the energy transfer introduced by the nondiffusive transport term
proportional to $\Lambda_{ik}$ responsible for maintaining the
differential rotation. The latter is in general the dominant term.
The importance of the term ${Q}$ for a stationary solution 
becomes more apparent if we transform the entropy equation to an 
equation of the quantity $\rh_0 T_0 s_1$, which better represents the
energy perturbation associated with the entropy perturbation. In the case of
a stationary solution, we have $\divv(\rh_0\mbox{\boldmath$v$})=0$ and we 
can rewrite the 
entropy equation Eq. (\ref{entr}) in the form (assuming 
$\vert\delta\vert=\vert\nabla-\nabla_{\rm ad}\vert\ll 1$)
\begin{eqnarray} 
    \divv\left(\mbox{\boldmath$v$}\,\rh_0 T_0 s_1
    -\kappa_t\rh_0 T_0\grad s_1\right)&=&\nonumber\\
    (\gamma-1)\left[\frac{\rh_0 T_0}{p_0}{Q}
      -v_r\frac{\rh_0 T_0 s_1}{\gamma H_p}\right]
    &+&v_r\gamma\delta\frac{\rh_0 T_0}{H_p}\;. \label{enthalpy}
\end{eqnarray}
The terms that appear in this equation are the divergence of the energy flux
(left-hand side), a source term that considers the viscous heating and the 
buoyancy work (first term, right-hand side), and the source term that arises 
from the nonadiabatic stratification (second term, right-hand side). 
The first term redistributes $\rh_0 T_0 s_1$ but does not provide a net 
source, since the flux across the 
boundaries vanishes with the boundary conditions we use. The same 
applies to the last term in the case of a stationary solution, since the
horizontal mean of the radial mass flux $v_r\rh_0$ has to vanish and 
$\delta$ and $g$ show no latitudinal dependence. The only net source
is the second term, which contains the Reynolds stress and
buoyancy work. In the case of a stationary solution the volume integral
of this term has to vanish, which means that the energy extracted through
the $\Lambda$-effect from the reservoir of internal energy (through feedback
on convective motions) returns through viscous heating and work of the 
meridional flow against the buoyant force. 

\subsection{Background stratification}
For the background $\rh_0$, $p_0$, and $T_0$ we use an adiabatic hydrostatic
stratification assuming an $\sim r^{-2}$ dependence of the gravitational
acceleration given by
\begin{eqnarray}
  T_0(r)&=&T_{\rm bc}\left[1+\frac{\gamma-1}{\gamma}\frac{r_{\rm bc}}
    {H_{\rm bc}}\left(\frac{r_{\rm bc}}{r}-1\right)\right]\;,\label{tref}\\
  p_0(r)&=&p_{\rm bc}\left[1+\frac{\gamma-1}{\gamma}\frac{r_{\rm bc}}
    {H_{\rm bc}}\left(\frac{r_{\rm bc}}{r}-1\right)\right]
       ^{\gamma/(\gamma-1)}\;,\label{pref}\\
  \rh_0(r)&=&\rh_{\rm bc}\left[1+\frac{\gamma-1}{\gamma}\frac{r_{\rm bc}}
    {H_{\rm bc}}\left(\frac{r_{\rm bc}}{r}-1\right)\right]
       ^{1/(\gamma-1)}\;,\label{rref}\\
  g(r)&=&g_{\rm bc}\left(\frac{r}{r_{\rm bc}}\right)^{-2}\;,
\end{eqnarray}
where $T_{\rm bc}$,  $p_{\rm bc}$, and  $\rh_{\rm bc}$ denote the values of
temperature, pressure and density at the base of the convection zone 
$r=r_{\rm bc}$. Here $H_{\rm bc}=p_{\rm bc}/(\rh_{\rm bc}\,g_{\rm bc})$ is the
pressure scale height and $g_{\rm bc}$ the value of the gravity at 
$r_{\rm bc}$. In the following we use $r_{\rm bc}=0.71\,R_{\odot}$
$p_{\rm bc}=6\times 10^{12}\,\mbox{Pa}$, 
$\rh_{\rm bc}=200\,\mbox{kg}\,\mbox{m}^{-3}$,
$g_{\rm bc}=520\,\mbox{m}\,\mbox{s}^{-2}$, and 
$R_{\odot}=7\times 10^8\,\mbox{m}$, which
results in $H_{\rm bc}= 0.0825\,R_{\odot}$.

\subsection{Superadiabaticity profile}
\label{superad}
For the superadiabaticity $\delta$ we assume the following profile:
\begin{equation}
  \delta=\delta_{\rm conv}+\frac{1}{2}(\delta_{\rm os}-\delta_{\rm conv})
  \left(1-\tanh\left(\frac{r-r_{\rm tran}}{d_{\rm tran}}\right)\right)\;,
  \label{delta}
\end{equation} 
where
\begin{equation}
  \delta_{\rm conv}=\delta_{\rm top}\exp(\frac{r-r_{\rm max}}{d_{\rm top}})
  +\delta_{\rm cz}\frac{r-r_{\rm sub}}{r_{\rm max}-r_{\rm sub}}\;.
\end{equation}

\begin{figure}
  \resizebox{\hsize}{!}{\includegraphics{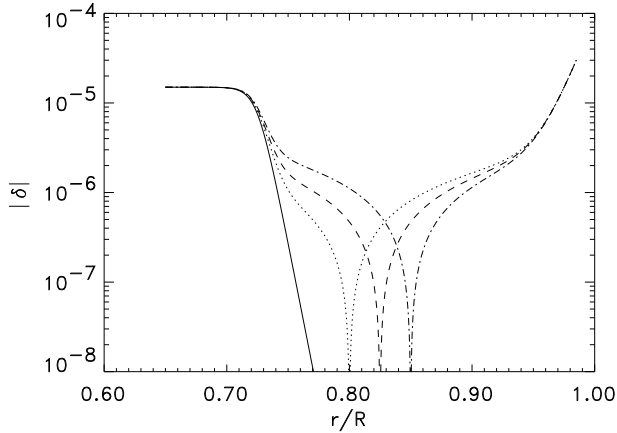}}
  \caption{Profile of superadiabaticity used in the models with adiabatic
    convection zone (solid line), model 8 (dotted line), model 9 (dashed line),
    and model 10 (dashed-dotted line). In all cases we have 
    $\delta_{\rm os}=-1.5\times 10^{-5}$, $r_{\rm tran}=0.725\,R_{\odot}$, and
    $d_{\rm tran}=0.0125\,R_{\odot}$. Common parameters of the models 8 to
    10 are $d_{\rm top}=0.0125\,R_{\odot}$, $\delta_{\rm cz}=3\times 10^{-6}$, 
    and $\delta_{\rm top}=3\times 10^{-5}$. Case 8 shows a profile with
    $r_{\rm sub}=0.8\,R_{\odot}$, case 9 with
    $r_{\rm sub}=0.825\,R_{\odot}$, and case 10 with 
    $r_{\rm sub}=0.85\,R_{\odot}$.
    Shown on the vertical axis is $\vert\delta\vert$ on a logarithmic scale. 
    The singularities at 
    $r=0.8$, $0.825$, and $0.85\,R_{\odot}$ indicate 
    where $\delta$ changes sign in the models with a nonadiabatic convection
    zone.
  }  
  \label{f1}
\end{figure}

Here $\delta_{\rm top}$, $\delta_{\rm cz}$, and $\delta_{\rm os}$ denote the
values of superadiabaticity at the top of the domain $r=r_{\rm max}$, 
in the bulk of the convection zone, and in the overshoot region, respectively. 
In addition $r_{\rm sub}$ denotes the radius at which the stratification 
within the convection zone turns weakly subadiabatic (because of nonlocal
effects), and $r_{\rm tran}$ denotes the radius of transition towards stronger 
subadiabatic stratification (the overshoot region). The parameters 
$d_{\rm top}$ and $d_{\rm tran}$ determine
the steepness of the transition toward large superadiabaticities at the top of
the domain and towards the overshoot region, respectively.
Fig. \ref{f1} shows the profiles of the superadiabaticity we use later
in our models.  

Nonlocal mixing-length models show a transition from subadiabatic to 
superadiabatic values typically between $r=0.75$ and 
$0.8\,R_{\odot}$ depending on the assumed mixing-length parameter 
\citep{Pidatella:Stix:1986,Skaley:Stix:1991}. 
For larger degrees of non-locality the subadiabatic fraction of the convection 
zone could be even larger \citep{Spruit:1997,Rempel:2004}. Mixing-length models
predict a variation of $\delta$ within the convection zone by several orders of
magnitude; however, most of this variation occurs
very close to the surface layers, which are difficult to resolve in a
mean field model. Since our model captures only the large-scale flows, we also 
have to make sure that the Rayleigh number stays sub critical for convection 
within the domain for the thermal conductivity and viscosity we use. 

In this investigation we keep the following parameters fixed:
$d_{\rm top}=d_{\rm tran}=0.0125\,R_{\odot}$ and 
$r_{\rm tran}=0.725\,R_{\odot}$. Choosing a value of 
$\delta_{\rm top}=3\times 10^{-5}$ we have to use a thermal diffusivity of 
around $5\times 10^9\,\mbox{m}^2\,\mbox{s}^{-1}$. Much larger 
values of $\delta$ would require unreasonably large values of the turbulent 
heat conductivity.
Below $r=r_{\rm tran}$ we assume overshoot type values of the 
superadiabaticity, in the range $-10^{-5}$ to $ -10^{-4}$.
We extend our domain down to $r=0.65\,R_{\odot}$, which also includes
the radiative interior with values of $\delta\sim -0.1$. Since our 
diffusivity profile (see next
subsection) drops significantly below $r_{\rm bc}$, the entropy perturbation
generated at the lower boundary does not influence the values within the
convection zone and has therefore no influence on the differential rotation
profile. More important for this model is the overlap between the
subadiabatic region and the thermal diffusivity profile, since this determines
to what extent an entropy perturbation can spread from the subadiabatic region
into the convection zone. Since the representation of a strongly subadiabatic 
layer would cause numerical difficulties (see section \ref{numerics} for 
further details) we decided to exclude this layer and use overshoot values for
$\delta$ down to $0.65\,R_{\odot}$.
The effective thickness of the overshoot is determined
by the overlap with the thermal conductivity profile, which we define
in the next subsection. Since the heat conductivity is basically zero below
$r_{\rm bc}=0.71\,R_{\odot}$, the effective thickness of the overshoot region 
is about $2\,d_{\rm tran}\approx 20\,\mbox{Mm}$, which is in the
range of predictions of nonlocal mixing-length models
\citep{Pidatella:Stix:1986,Skaley:Stix:1991}.

Since all these values are small compared with $\nabla_{\rm ad}$, we still can 
use the adiabatic reference state for pressure, density, and temperature as 
given above in Eqs. (\ref{tref}) to (\ref{rref}).

\subsection{Diffusivity profiles}
\label{diffprof}
For the turbulent viscosity and thermal conductivity we assume constant
values within the convection zone and the radiative zone with a transition
smoothed by a hyperbolic tangent function. We assume that the diffusivities 
only depend on the radial coordinate: 
\begin{eqnarray}
  \nu_t&=&\frac{\nu_0}{2}\left[1+\tanh\left(\frac{r-r_{\rm tran}+\Delta}
    {d_{\kappa\nu}}\right)\right]f_c(r)\;,
  \label{visc}\\
  \kappa_t&=&\frac{\kappa_0}{2}\left[1+\tanh\left(\frac{r-r_{\rm tran}+\Delta}
    {d_{\kappa\nu}}\right)\right]f_c(r)
  \label{cond}\;,
\end{eqnarray}
with
\begin{eqnarray}
  f_c(r)&=&\frac{1}{2}\left[1+\tanh\left(\frac{r-r_{\rm bc}}{d_{\rm bc}}
    \right)\right]\;,\\
  \Delta&=&d_{\kappa\nu}\tanh^{-1}\left(2\alpha_{\kappa\nu}-1\right)\;,
\end{eqnarray} 
where $\nu_0$ and $\kappa_0$ denote the values of the turbulent diffusivities
within the convection zone and $\alpha_{\kappa\nu}$ specifies the values of 
the 
turbulent diffusivities at $r=r_{\rm tran}$ ($\alpha_{\kappa\nu}\nu_0$, 
$\alpha_{\kappa\nu}\kappa_0$), where 
the transition to the overshoot region takes place in our model. The profile 
function $f_c$ ensures that the diffusivities drop significantly toward the
radiative interior at $r_{\rm bc}$. For the width of this transition we use
$d_{\rm bc}=0.0125\,R_{\odot}$. Since both diffusivities are of 
turbulent origin, we use the same radial profile.

In this model we do not include a sophisticated theory for the tachocline 
but force a tachocline through a uniform rotation boundary at 
$r=0.65\,R_{\odot}$. As a consequence the tachocline is a viscous shear layer 
between the convection zone and the lower boundary condition. We therefore
have to maintain a sufficient amount of viscosity in the radiative  interior
to allow for the formation of this shear layer in a reasonable amount of time.
In the following we use the profile defined by Eq. (\ref{visc}) for the 
computation of the turbulent angular momentum transport  (last term in
Eq. [\ref{stresstensor}]) but set the viscosity used for the diffusive
terms of the Reynolds stress (first two terms in Eq. [\ref{stresstensor}])
to $2\%$ of the convection zone values. For the heat conductivity we use
in the radiative interior $0.2\%$ of the convection zone values.

\subsection{Parameterization of turbulent angular momentum transport}
The terms relevant for the differential rotation are:
\begin{eqnarray}
  \Lambda_{r\phi}&=&\Lambda_{\phi r} = 
     +{L}(r,\theta)\,\cos(\theta+\lambda(r,\theta))\label{lambda_a}\;,\\
  \Lambda_{\theta\phi}&=&\Lambda_{\phi\theta} = 
     -{L}(r,\theta)\,\sin(\theta+\lambda(r,\theta))\label{lambda_b}\;,
\end{eqnarray}
where ${L}(r,\,\theta)$ denotes the amplitude of the angular momentum
flux, whereas $\lambda(r,\,\theta)$ describes the inclination of the
flux vector with respect to the axis of rotation. 

Symmetry considerations require that $\Lambda_{r\phi}$ is symmetric and 
$\Lambda_{\theta\phi}$ is antisymmetric across the equator. To satisfy this
constraint, $\lambda(r,\,\theta)$ and ${L}(r,\,\theta)$ need to be 
antisymmetric functions with respect to the equator. Since we solve our model 
only in the northern hemisphere, we specify in the following discussion 
always values for $\lambda$ and ${L}$ in the northern hemisphere. 
To maintain the proper symmetry for a full-sphere simulation,
values for the southern hemisphere would have to be chosen by reflection
across the equator.

The setting 
${L}>0$, $\lambda=0$ corresponds to a flux directed downward and
parallel to the axis of rotation; ${L}>0$, $\lambda=-\theta$ 
to a radially inward flux, and ${L}>0$, $\lambda=\pi/2-\theta$ to
an equatorward flux. The setting ${L}\sim \sin\theta\cos\theta$ 
and $\lambda=0$ recovers the limit of fast rotation found for the
$\Lambda$-effect by \citet{Kitchatinov:Ruediger:1995}.  

In the following discussion we will for the amplitude of the
angular momentum flux the expression
\begin{eqnarray}
  f(r,\theta)&=&(\sin\theta)^n\cos\theta
  \tanh\left(\frac{r_{\rm max}-r}{d}\right)\nonumber\;,\\
  {L}(r,\theta)&=&\Lambda_0\Omega_0\frac{f(r,\theta)}
	  {\max\vert f(r,\theta)\vert}\label{amflux}\;.
\end{eqnarray}
The full radial dependence of the angular momentum flux is obtained
by multiplication by $\nu_t \rh_0$. The angular momentum flux drops below 
$r_{\rm tran}$ because of the drop in turbulent viscosity. In most of the 
following discussion we require a vanishing angular momentum flux at the
top boundary as expressed in Eq. (\ref{amflux}), where we use a value of 
$d=0.025\,R_{\odot}$ for the transition layer at the top. The exponent $n$
determines the latitude at which the flux peaks. For $\lambda>0$, the value 
of $n$ needs to be larger than $2$ to ensure the regularity of the divergence 
of the Reynolds stress close to the pole. For the direction of
the flux determined by $\lambda$ we discuss two distinct cases:
$\lambda=15\degr$ (transport nearly parallel to axis of rotation) and 
$\lambda=90\degr-\theta$ (transport in latitude only) in order 
to evaluate the sensitivity of the model with respect to this 
parameterization. 

\subsection{Numerical procedure}
\label{numerics}
We are interested here in the stationary solution of Eqs. (\ref{dens})
to (\ref{entr}) for a given parameterization of the turbulent angular momentum
transport Eq. (\ref{lambda_a}), (\ref{lambda_b}). 
A very natural way to relax the system is to use the temporal evolution; 
however, because of the low Mach number of the 
expected flows, a direct compressible simulation is problematic. 
For an expected meridional flow velocity of a few 
$\mbox{m}\,\mbox{s}^{-1}$ the Mach number is around $10^{-5}$ 
(because of axisymmetry, the much faster differential
rotation flow does not enter the time step limit). Without leaving the regime 
of highly subsonic flows and therefore without changing the physical properties
of the solution, it is possible to speed up the relaxation process 
significantly by increasing the base rotation rate $\Omega_0$. 

Using the following transformation for the independent parameters
of the equations,
\begin{eqnarray}
  \Omega_0       &\longrightarrow& \zeta \Omega_0\;,\nonumber\\
  \nu_t          &\longrightarrow& \zeta\nu_t\;,\nonumber\\
  \kappa_t       &\longrightarrow& \zeta\zeta\kappa_t\;,\nonumber\\
  \delta         &\longrightarrow& \zeta^2\delta\;,\nonumber\\
  t              &\longrightarrow& \zeta^{-1}t\;,\nonumber
\end{eqnarray}
and the following transformation for variables,
\begin{eqnarray}
  v_r\longrightarrow\zeta v_r,\; & v_{\theta}\longrightarrow\zeta v_{\theta}
  ,\;&\Omega_1 \longrightarrow \zeta\Omega_1\;,\nonumber\\
  p_1\longrightarrow\zeta^2 p_1,\;&\rh_1\longrightarrow\zeta^2 \rh_1,\;&
  s_1\longrightarrow\zeta^2 s_1\;,\nonumber\\
\end{eqnarray}

Eqs. (\ref{vrad}) - (\ref{entr}) remain unchanged, meaning that if
$\{\rh_1,\;v_r,\;v_{\theta},\;\Omega_1,\;s_1\}$ is a solution for the 
parameters
$\{\Omega_0,\;\nu_t,\;\kappa_t,\;\delta\}$ then 
$\{\zeta^2\rh_1,\;\zeta v_r,\;\zeta v_{\theta},\;\zeta \Omega_1,\;\zeta^2 s_1\}$
is a solution for the parameters  
$\{\zeta \Omega_0,\;\zeta \nu_t,\;\zeta \kappa_t,\;\zeta^2\delta\}$. However
this transformation changes the equation of continuity to:
\begin{equation}
  \frac{\partial \rh_1}{\partial t}+\frac{1}{\zeta^2}
  \divv(\rh_0\,\mbox{\boldmath$v$})=0\;,
\end{equation}
meaning that the time evolution is changed, but the stationary solution 
remains unchanged. The pre factor $\zeta^{-2}$ in the equation of continuity
corresponds to the increase of the Mach number of the flow by a factor of 
$\zeta$. In the following we use a value of $\zeta=100$, which
corresponds to an increase of the Mach number of the meridional flow in the
bulk of the convection zone
from $10^{-5}$ to $10^{-3}$. Therefore even the time evolution is only
marginally affected, since the solution stays in the regime of highly
subsonic flows. We computed solutions with different values of $\zeta$ in 
order to quantify the influence of this transformation. We found differences 
between a solution computed with $\zeta=10$ and  $100$ on the order 
of a few percent (mainly in the magnitude of the differential rotation close
to the pole).

This approach is very similar to simulations of 
rising magnetic flux tubes in the solar convection zone, which were made
fully compressible by assuming a value of $\beta=p_{\rm gas}/p_{\rm mag}$
on the order of $100$ instead of $10^5$ to overcome a severe time step
constraint. As long as the relevant flow velocities remain sufficiently
subsonic, the results are not affected significantly.

The only drawback of this approach is that it is impossible to 
represent large values of the subadiabaticity as found in the radiative 
interior of the Sun. Because of the scaling of $\delta$ with $\zeta^2$,
a value of $\delta=-0.1$ would correspond to $\delta=-1000$ in a solution
with a base rotation increased by a factor of $100$, which is physically
impossible. However, overshoot-like values of $\delta\sim -10^{-5}$ can
be treated without any problem. As we explained earlier in subsection
\ref{superad} and \ref{diffprof}, the radiative interior is not of great 
importance, since entropy perturbations created there cannot influence 
the convection zone we are primarily interested in.

Since our model also includes a significant time step constraint because of the
large turbulent diffusivities in the convection zone, an anelastic approach
would not provide much advantage unless all diffusivities were treated 
implicitly. We therefore decided to solve the equations with a faster 
explicit scheme using the procedure outlined above. We tested an anelastic 
version of the code and found convergence problems of the pressure solver 
related to the uniform rotation boundary that we impose at $r=0.65\,R_{\odot}$. 
For a different choice of boundary conditions we found very good agreement 
between the anelastic version and a solution computed with $\zeta=100$ as 
described above. 

We solve Eqs. (\ref{dens}) - (\ref{entr}) with a MacCormack scheme using
alternating upwind and downwind differencing, which is second order in space 
and time. The computational domain  extends in latitude from equator to pole 
and in radius from $r=0.65$ to $0.985\,R_{\odot}$. We use the 
appropriate symmetry boundary conditions at equator and  pole and closed 
boundaries in radius. At the bottom boundary ($r=0.65\,R_{\odot}$) we 
enforce a uniform rotation; the top boundary is stress-free for the angular
velocity (${R}_{r\phi}=0$). At both radial boundaries we use 
stress-free boundary conditions for the velocity and set the derivative
of $s_1$ to zero. Since we focus in this study on the large-scale flow 
fields, a moderate resolution of around $108$ grid points in radius and $72$
grid points in latitude is sufficient.
We tested our code by reproducing the result presented by 
\citet[][their Fig. 1 ]{Ruediger:etal:1998}.

\begin{figure*}
  \resizebox{\hsize}{!}{\includegraphics{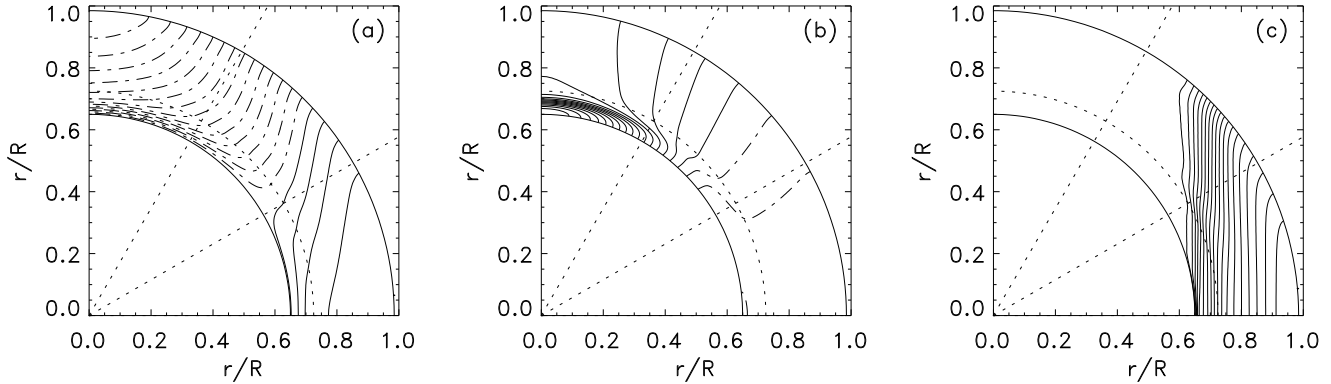}}
  \caption{Contours of (a) differential rotation and (b) entropy perturbation
    for case 1. Solid lines indicate positive values. The entropy 
    perturbation that originates in the subadiabatic tachocline and spreads 
    because of thermal
    conductivity into the convection zone prevents the Taylor-Proudman state
    (contours parallel to axis of rotation) for the differential rotation
    from developing.
    (c) Contours of differential rotation for case 3. This case is similar
    to case 1, except that $s_1=0$. As a consequence the contour lines
    of constant $\Omega$ are aligned with the axis of rotation. 
  }  
  \label{f2}
\end{figure*}

\section{Results}

\begin{deluxetable}{ccccccccc}
  \tablecaption{\label{tab1}Significant parameters of simplified model}
  \tablehead{ \colhead{case} & \colhead{$\lambda$} & \colhead{$n$} & 
    \colhead{$d_{\kappa\nu}$} & \colhead{$\alpha_{\kappa\nu}$} & 
    \colhead{$\delta_{\rm os}$} & \colhead{$r_{\rm sub}$} }
  \startdata
  1 & $15\degr$ & $2$ & $0.025$ & $0.1$  & $-1.5\times 10^{-5}$ &\\
  2 & $90\degr-\theta$ & $2$ & $0.025$ & $0.1$  & $-1.5\times 10^{-5}$ &\\
  3 & $15\degr$ & $2$ & $0.025$ & $0.1$  & $0$                 &\\
  4 & $15\degr$ & $4$ & $0.025$ & $0.1$  & $-1.5\times 10^{-5}$ &\\
  5 & $15\degr$ & $2$ & $0.025$ & $0.1$  & $-3\times 10^{-5}$   &\\
  6 & $15\degr$ & $2$ & $0.05$  & $0.1$  & $-1.5\times 10^{-5}$ &\\
  7 & $15\degr$ & $2$ & $0.025$ & $0.025$& $-1.5\times 10^{-5}$ &\\
  \\[0.05cm]
  8 & $15\degr$ & $2$ & $0.025$ & $0.1$  & $-1.5\times 10^{-5}$ & $0.8$\\
  9 & $15\degr$ & $2$ & $0.025$ & $0.1$  & $-1.5\times 10^{-5}$ & $0.825$\\
  10 & $15\degr$ & $2$ & $0.025$ & $0.1$  & $-1.5\times 10^{-5}$ & $0.85$\\
  \enddata
  \tablecomments{Cases 1 - 7 have adiabatic convection zones and cases
    8 - 10 have nonadiabatic convection zones with, the common parameters
    $\delta_{\rm cz}=3\times 10^{-6}$ and $\delta_{\rm top}=3\times 10^{-5}$,
    but different values of $r_{\rm sub}$, as shown in the table.}
\end{deluxetable}
 
Since our simplified model contains parameterizations of crucial processes,
we have to evaluate carefully the dependence on particular choices of 
these parameters. In Table \ref{tab1} we have summarized the parameters that
we discuss in this section. There are additional model parameters, which
do not have a significant influence on the solution. These are $r_{\rm tran}$,
$\Lambda_0$, $\nu_0$, and $\kappa_0$. Here $r_{\rm tran}$ specifies where the 
superadiabaticity turns from convection zone to overshoot values.
In the following we use 
$r_{\rm tran}=0.725\,R_{\odot}$, which is a reasonable choice for the Sun. 
As long as $r_{\rm tran}>r_{\rm bc}+d_{\rm bc}$ the influence 
on the solution is small, since it is more the profile of $\nu_t$ and 
$\kappa_t$ in relation to $r_{\rm tran}$ that matters. We have therefore 
introduced in Eqs. (\ref{visc}) and (\ref{cond}) the parameter 
$\alpha_{\kappa\nu}$, which specifies the profile relative to $r_{\rm tran}$.
Similarly $\Lambda_0$ specifies the magnitude of the nondiffusive 
Reynolds-stress
($\Lambda$-effect). Typical values for $\Lambda_0$ are on the order of unity.
Except for case 2, in which we use $\Lambda_0=0.4$, in the following we use a 
value of $\Lambda_0=0.8$. We have chosen $\Lambda_0$ such that the magnitude 
of the differential rotation is close to solar-like. 

For the diffusivities $\nu_0$ and $\kappa_0$  we assume in the 
following discussion  $\nu_0=\kappa_0=5\times 10^8\,\mbox{m}^2\,\mbox{s}^{-1}$.
For cases with a superadiabatic convection zone we have to increase the
value of $\kappa_0$ to $5\times 10^9\,\mbox{m}^2\,\mbox{s}^{-1}$ above
$r_{\rm sub}$ in order to avoid convective instability.

\subsection{General solution properties: differential rotation}
The key ingredient in this model is the inclusion of a subadiabatic tachocline
beneath the convection zone, which is enforced in this model through the 
uniform rotation lower boundary condition. Within the subadiabatic region 
($r\lesssim0.725\,R_{\odot}$) the differential rotation is balanced by a 
latitudinal entropy gradient. Taking the
curl of the meridional momentum equation yields for the $\phi$ component
of the vorticity under the assumption of small deviations from adiabaticity
($\vert \nabla-\nabla_{\rm ad}\vert \ll 1$)
\begin{equation}
  \frac{\partial \omega_{\phi}}{\partial t}=[\ldots]+
  r\sin\theta\frac{\partial \Omega^2}{\partial z}
  -\frac{g}{\gamma r}\frac{\partial s_1}{\partial \theta} 
  \label{balance}
\end{equation} 
with $\Omega=\Omega_0+\Omega_1$; the bracket denotes viscous terms and
vorticity transport terms, which are not important for the following 
discussion.

Starting initially with $s_1=0$, the turbulent angular momentum transport
leads to a negative values of $\partial \Omega^2/\partial z$ in high latitudes,
which enforces a negative value of $\omega_{\phi}$. A negative value of
$\omega_{\phi}$ corresponds to a counterclockwise meridional flow
in the tachocline, which shows a negative radial velocity at high latitudes
and a positive velocity at low latitudes. Because of the subadiabatic 
stratification,
this results in a positive entropy perturbation in high latitudes and a 
negative entropy perturbation in low latitudes, as shown in Fig. \ref{f2} b).
Since the resulting negative value of $\partial s_1/\partial \theta$ can
compensate for the also negative value of $\partial \Omega^2/\partial z$,
an equilibrium is reached finally. An additional source for the entropy
perturbations comes from the meridional flow driven in the convection zone
and penetrating to some extent into the subadiabatic overshoot region.
For the parameters used in this model both effects are of roughly the same 
order of magnitude. In our model most of the tachocline
shear is located below the base of the convection zone, whereas helioseismic
inversions find more overlap between the tachocline and the convection zone
\citep{Charbonneau:etal:1999}. Our model has therefore most probably the 
tendency to underestimate the entropy perturbation in the overshoot region 
caused by the value of $\partial \Omega^2/\partial z$.

Because of the turbulent thermal heat conductivity this entropy perturbation 
can spread into the convection zone and therefore also balance there
a differential rotation that deviates from the Taylor-Proudman state
with $\Omega$-contours parallel to the axis of rotation. 
We want to emphasize that the total entropy $s_0+s_1$ in the overshoot region 
is still smaller than in the convection zone. The physical reason for
this spread is that the convection tries
to maintain the same radial entropy gradient at all latitudes
(if we do not consider possible rotational anisotropy). Since the overshoot
region provides the entropy boundary condition for the convection zone,
a latitudinal variation of entropy in the overshoot region is transported
by convection into the convection zone. \citet{Stix:1981} computed response
functions for the temperature, velocity, and flux perturbations within in the
framework of the mixing-length approach and found that the screening effect
of temperature perturbations is very weak, meaning that temperature (entropy)
perturbations at the base of the convection zone should be transmitted
through the entire convection zone.  

The magnitude
of the entropy perturbation in the convection zone depends on the overlap 
between the thermal conductivity profile and the subadiabaticity profile.
For most models we use a parameter of $\alpha_{\kappa\nu}=0.1$, which means
that the thermal diffusivity drops to $10\%$ of its convection zone values at
$r=r_{\rm tran}=0.725\,R_{\odot}$, but smaller values 
($\alpha_{\kappa\nu}=0.025$) also work if the value of $\delta_{\rm os}$ is 
slightly increased (see cases 5 and 7). The thermal heat diffusivity 
in a nonlocal mixing-length model, $\sim v_{\rm conv} H_p$, would lead to
larger values within the overshoot region because of the large value of
the pressure scale height and still significant velocities 
$\sim 10\,\mbox{m}\,\mbox{s}^{-1}$
in the overshoot region.  

The temperature perturbation associated with the entropy perturbation shows
a magnitude of about $5\mbox{K}$ throughout the convection zone for most of
our models. Whereas the entropy perturbation drops monotonically from pole
to equator, the temperature perturbation close to the surface reaches a 
minimum around mid latitudes followed by a slight increase toward the equator.
At the base of the convection zone the temperature also shows a monotonic
decrease from the pole toward the equator. The different behavior of entropy
and 
temperature is due to the pressure perturbation within the convection zone,
which also contains an adiabatic contribution (first term on right-hand side
of Eq. [\ref{pressure}]). A similar pattern was also found by 
\citet{Brun:Toomre:2002} in three-dimensional simulations; however, the 
physical reason might be different.

In Fig. \ref{f2} we show the contours of $\Omega$ and the related
entropy perturbation $s_1$ for case 1. Since the entropy perturbation
is concentrated in higher latitudes (where $\partial \Omega^2/\partial z$ also
peaks), the deviations from the Taylor-Proudman
state are largest in high latitudes. Whereas the differential rotation
shows close to the pole $\Omega$-contours perpendicular to the axis of
rotation, the $\Omega$-contours are more aligned with the axis of rotation
close to the equator. 

In Fig. \ref{f2} c) we show for reference purposes a solution with
$s_1=0$ but, apart from that, exactly the same parameters as case 1. Without 
the effect of the subadiabatic tachocline the solution shows differential 
rotation
with cylindrical $\Omega$ contours, even though we still impose the uniform
rotation boundary condition at $r=0.65\,R_{\odot}$.

\begin{figure*}
  \resizebox{\hsize}{!}{\includegraphics{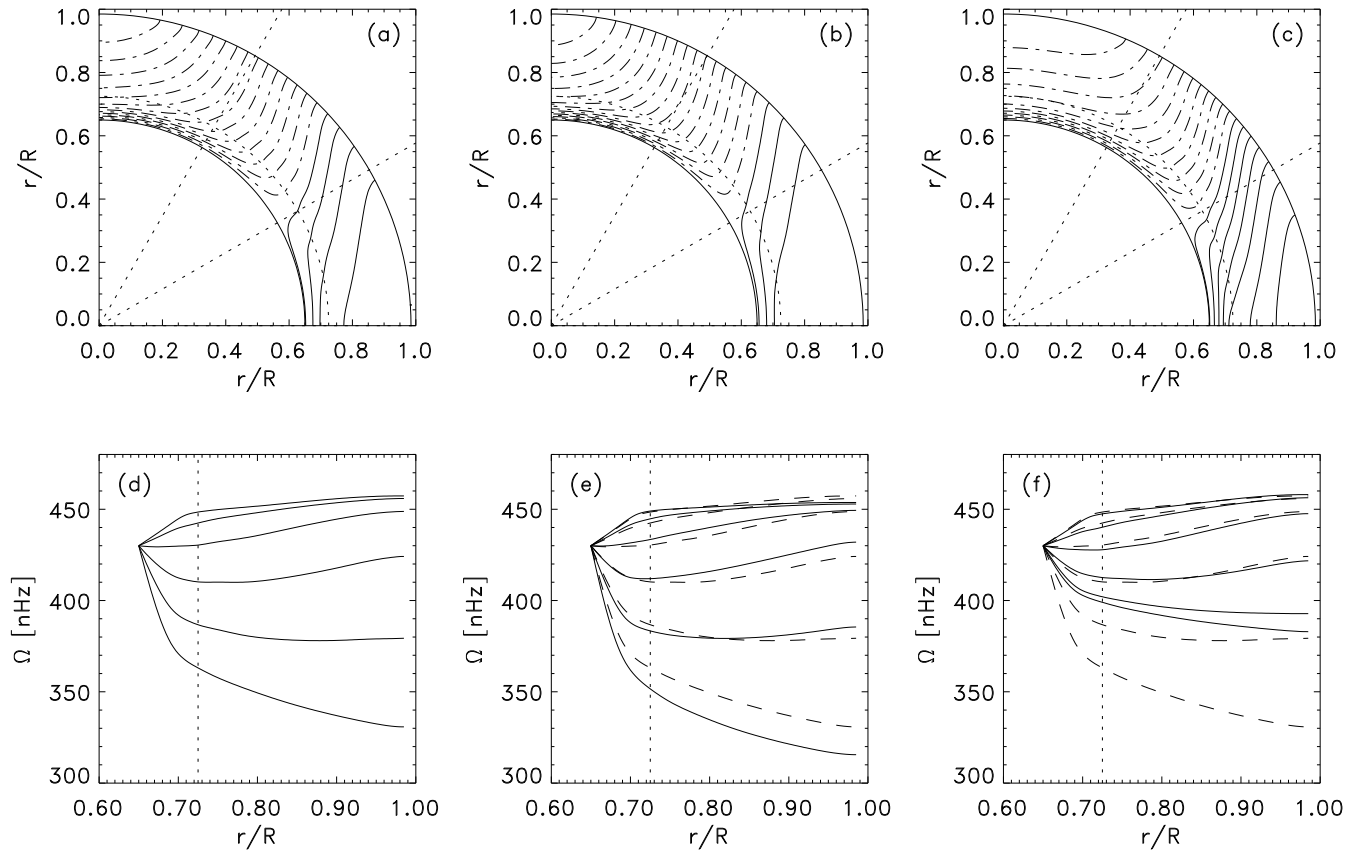}}
  \caption{Influence of different parameterizations of the turbulent angular 
    momentum transport on differential rotation.
    (a - c) Contours of the differential rotation, where
    solid lines indicate regions rotating faster than the core. (d - f)
    Profile of the differential rotation as a function of radius for
    the latitudes $0\degr$, $15\degr$, $30\degr$, $45\degr$, $60\degr$, 
    and $90\degr$.
    Differential rotation is shown for (a, d) case 1 ($\lambda=15\degr$), 
    (b, e) case 2 ($\lambda=90\degr-\theta$), and (c, f) case 4 
    ($\lambda=15\degr$ and $n=4$). The profile of the differential rotation is
    rather insensitive to changes in the Reynolds stress; however, 
    the amplitude changes, especially in high latitudes.
  }
  \label{f3}
\end{figure*}

\begin{figure*}
  \resizebox{\hsize}{!}{\includegraphics{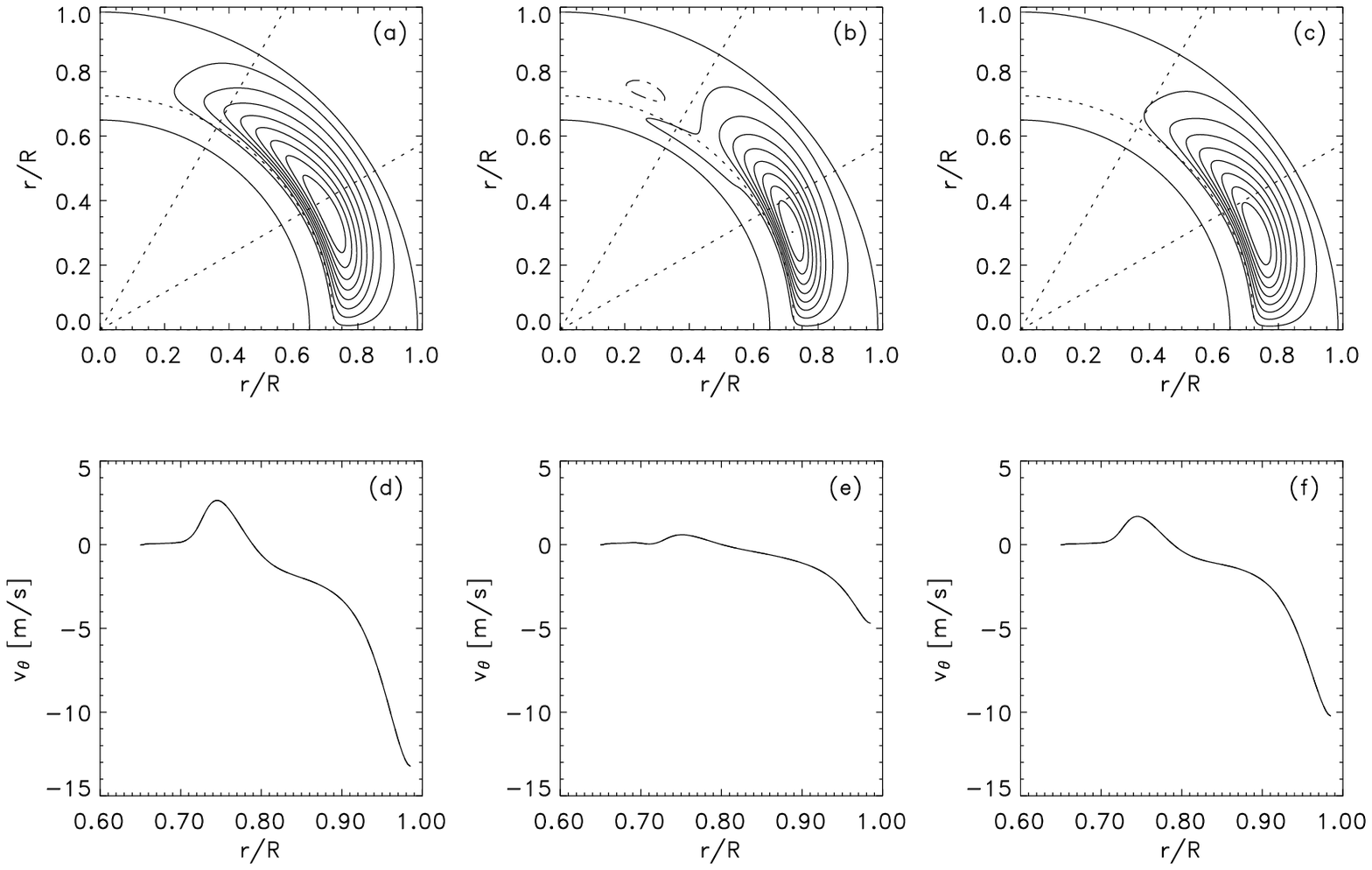}}
  \caption{Influence of different parameterizations of the turbulent 
    angular momentum transport on meridional flow. 
    (a - c) Stream function, where solid lines
    indicate counterclockwise flows. (d - f) Radial 
    profile of $v_{\theta}$ at $45\degr$ latitude. Meridional flow is shown 
    for (a, d) case 1, (b, e) case 2, and (c, f) case 4.  Unlike 
    the differential rotation, the meridional flow is very sensitive to the 
    direction of the angular momentum transport (compare cases 1 and 2).
    In case 4 the confinement of the Reynolds stress to lower latitudes also
    confines the meridional flow to lower latitudes.
  }

  \label{f4}
\end{figure*}

\subsection{General solution properties: meridional flow}
Since the assumed turbulent angular momentum transport has in general the 
tendency to drive a differential rotation that differs from the profile that 
would balance the right-hand side of Eq.
(\ref{balance}), the differential rotation shows a small perturbation
around this state. As a consequence, the Coriolis force related to this 
perturbation cannot be precisely balanced by a combination of pressure 
gradient and buoyancy and therefore drives a meridional flow (examples of 
this flow can be seen in Fig. \ref{f4}). This flow 
grows until the additional angular momentum transport sufficiently limits the 
perturbation of $\Omega$ so that the remaining weak unbalanced Coriolis 
force can be balanced by viscous stress associated with the meridional flow.
For the parameterization of the Reynolds stress used in our model we typically
get a counterclockwise flow cell (equatorward at base and 
poleward at surface).
This is strongly related to the radial component of the turbulent angular 
momentum flux, which is assumed to be directed inward in most of our models.
In the absence of a meridional flow, it would increase the rotation rate at the
base and decrease the rotation rate at the surface, which leads to 
counterclockwise Coriolis forces.

The meridional flow typically closes above $r=0.71\,R_{\odot}$ for two reasons.
(1) The subadiabatic stratification suppresses radial motions. (2) The
Reynolds stress ($\sim \nu_t$) that indirectly drives a meridional flow 
through a change of $\Omega$ drops significantly below $r=0.71\,R_{\odot}$. 
The meridional flow will never vanish completely in the subadiabatic
region, since some radial motion is required to maintain the entropy
perturbation against diffusive decay. However, the flow velocities are 
very small compared with the values within the convection zone. For our
choice of a value of $\delta\sim -10^{-5}$, the value of $v_{\theta}$
is on the order of a few $\mbox{cm}\,\mbox{s}^{-1}$. For values
of $\delta\sim -10^{-1}$ as found in the radiative interior, these values
would drop by at least 3 orders of magnitude. The limits for the 
penetration of the meridional flow below the base of the convection zone 
are even more stringent in this model than the analysis of 
\citet{Gilman:Miesch:2004} showed for overshoot-like values of $\delta$.
The main reason for this is the inclusion of the feedback of the meridional
flow on the differential rotation through the transport of angular momentum.
Any significant equatorward flow in a region with low
turbulent diffusivity would cause a retrograde zonal flow that would
oppose the flow through the Coriolis force. Beside the effect of
subadiabaticity to suppress radial motions, the angular momentum conservation
in a low-diffusivity region yields an additional effect suppressing
latitudinal motions. Only if there is a process that disturbs an equilibrium
solution of Eq. (\ref{balance}) can there be a significant
meridional flow below the base of the convection zone.  
 
\subsection{Dependence on parameterization of angular momentum transport}
Figs. \ref{f3} and \ref{f4} show the sensitivity of the solution with respect
to different parameterizations of the turbulent angular momentum 
transport. In cases 1 and 2 the amplitude of the angular momentum flux
has the same profile in radius and latitude; however, the direction of the
flow is changed. In case 1 the angular momentum is transported almost parallel
to the axis of rotation with a $15\degr$ inclination angle; in case 2 the 
angular momentum flux has only an equatorward latitudinal component 
($\lambda=90\degr-\theta$). In both cases the
amplitude of the angular momentum flux, $\Lambda_0$, was adjusted such that 
the equatorial rotation rate is the same ($\Lambda_0=0.8$ in case 1 and
$\Lambda_0=0.4$ in case 2). The variation in the required amplitude
follows from the fact that an angular momentum transport perpendicular
to the axis of rotation is the most efficient way to speed up the rotation at
the equator, 
whereas a transport parallel to the axis of rotation would have no effect at 
all. A comparison of Fig. \ref{f3}, panel a) and b), shows that the profile of 
the differential rotation is not very sensitive to the change in the direction 
of the angular momentum. 

Fig. \ref{f3}, panels c) and f) show case 4, which is similar to case 1, but
uses a turbulent angular momentum transport confined closer to the equator
($n=4$ instead of $n=2$). Whereas the profile of the differential rotation
is only marginally changed, the magnitude of the differential rotation in high 
latitudes is reduced.
 
Unlike $\Omega$, the meridional flow is much more sensitive to the details
of the angular momentum transport. Fig. \ref{f4}, panels a) to c), show
the streamlines of the meridional flow, where solid lines indicate 
counterclockwise flows (poleward at the surface and equatorward at the base 
of the convection 
zone). Fig. \ref{f4}, panels d) to f), show the latitudinal component of the 
meridional flow velocity at $45\degr$ latitude. All cases show a dominant 
counterclockwise cell; however, each case shows different flow amplitudes. 
The cases 1 and 3 with a significant angular momentum flux along the axis of 
rotation show flow velocities by a factor of around $3$ larger, which is in 
part caused by the larger value of $\Lambda_0$ required in these cases 
to obtain the same
equatorial rotation rate. Case 2 with angular momentum flux in latitude only
also shows a weak reverse circulation cell above $60\degr$ latitude. The second
cell is driven by buoyancy resulting from the higher value of the entropy
close to the pole. This second cell is not visible in case 1, 
since the radial component of the angular momentum flux has a strong
tendency to drive a counterclockwise meridional flow. The radially inward flux
of angular momentum at high latitudes (because of the assumed $15\degr$ 
inclination with respect to the axis of rotation) increases the value
of $\partial \Omega^2/\partial z$ in high latitudes, which leads to a 
counterclockwise meridional flow according to Eq. (\ref{balance}). There is
also a very weak indication of this second cell in case 4, in which the larger
value of $n$ confines the meridional flow to lower latitudes.

The main reason for the different sensitivities of meridional flow and 
differential rotation with respect to changes in the Reynolds stress follows
from the fact that the differential rotation is mainly determined by the
balance expressed in Eq. (\ref{balance}). Since the entropy perturbation
is not directly affected by a changing Reynolds stress, the profile of
$\Omega$ also changes little. The meridional flow, on the other hand, results
from an imbalance of Eq. (\ref{balance}) and is therefore the result of 
a small difference between large forces. As consequence, the sensitivity of
the meridional flow to changes in the Reynolds stress is much larger.

We focus here on two distinct cases for the value of $\lambda$. We want to
mention that any choice for $\lambda$ $>0\degr$ and $<90\degr$ leads to 
differential 
rotation close to case 1 (large values of $\lambda$ are closer to the 
Taylor-Proudman state in low latitudes); however, the meridional flow changes 
significantly. Changing the value of $\lambda$ requires an adjustment of 
$\Lambda_0\sim 1/\sin\lambda$ in order to keep the magnitude of the
differential rotation fixed. As a consequence, models with smaller values
of $\lambda$ have larger meridional flow velocities than models with larger
values of $\lambda$. Using a value of $\lambda>45\degr$ also leads to a more 
complicated flow structure, which shows a clockwise flow pattern within most 
of the convection zone. Using values of $\lambda<15\degr$ leads to solutions 
very similar to case 01, but with a significantly larger amplitude of the 
meridional flow velocity.   

\begin{figure*}
  \resizebox{\hsize}{!}{\includegraphics{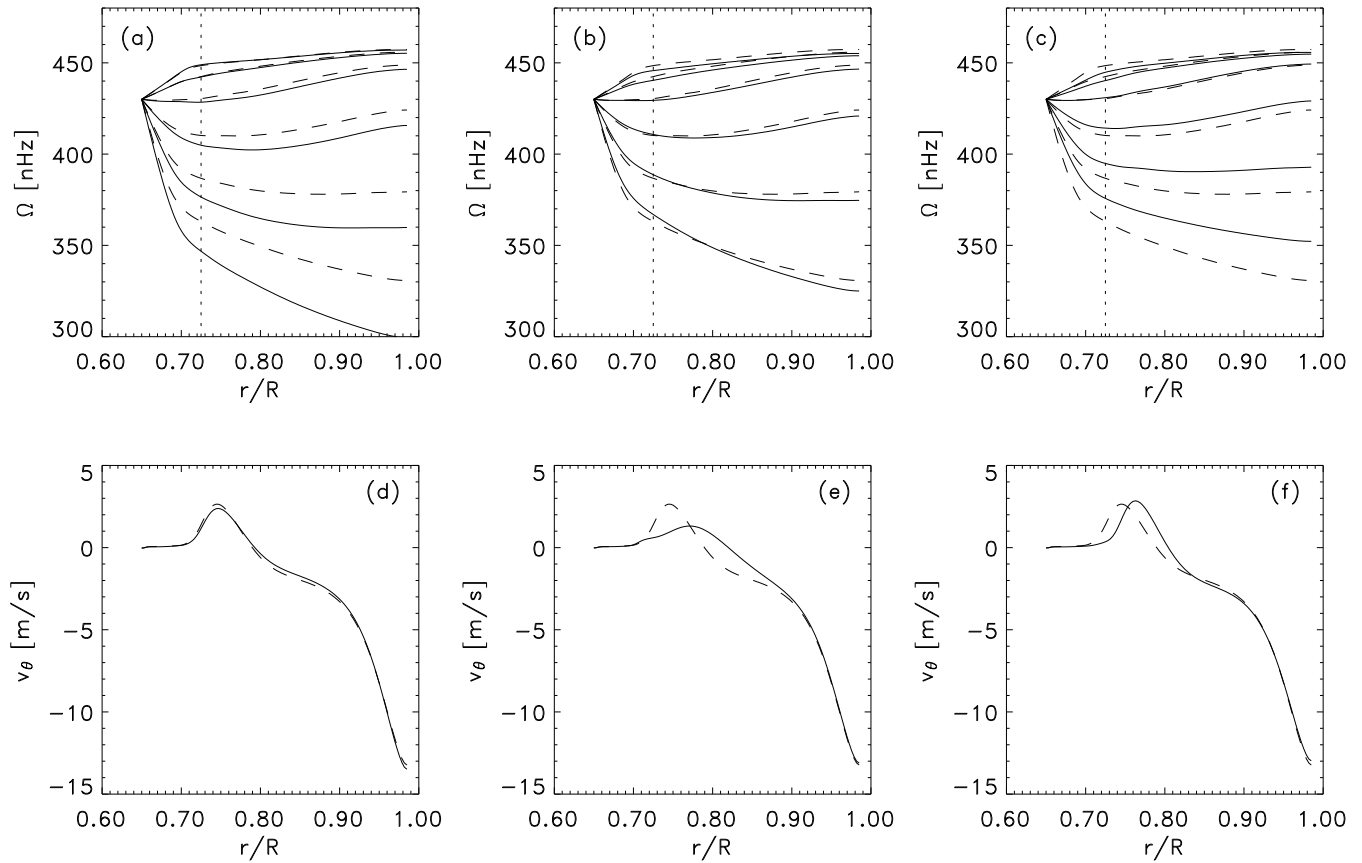}}
  \caption{Influence of subadiabaticity, viscosity, and conductivity on 
    solution. (a, d) Case 5 with a subadiabaticity increased
    by a factor of $2$, leading to larger equator-pole difference in $\Omega$
    but nearly no change in the meridional flow pattern. (b, e) Case 6 with
    an increased value of $d_{\kappa\nu}$, leading to significant reduction 
    in the meridional flow speed. (c, f) Case 7 with a decrease in 
    $\alpha_{\kappa\nu}$  
    Decreasing the overlap between the diffusivity profile and the 
    subadiabaticity profile is very similar to decreasing the value of
    $\delta$, except that the meridional return flow at the base
    of the convection zone is located at a different depth. In all
    panels we have shown case 1 as a dashed line for reference.
  }
  \label{f5}
\end{figure*}

\subsection{Dependence on subadiabaticity, viscosity, and thermal conductivity
profile}
Fig. \ref{f5} compares models with identical parameterization of the 
Reynolds stress but different parameters for the subadiabaticity and the 
profiles
of viscosity and heat conductivity. Panels a) and d) show how different
values of the subadiabaticity affect differential rotation and meridional
flow. In both panels the case 1 is shown as a reference (dashed line). 
Case 5 with a value of $\delta_{\rm bc}$ twice as large shows an increase of 
the
differential rotation  by about $20\%$. Whereas the differential rotation
remains unchanged at the equator, the polar values of $\Omega$ decrease 
significantly. On the other hand, the meridional flow is only marginally
affected. A larger value of $\delta_{\rm bc}$ leads to a
larger entropy perturbation, which can balance a larger differential rotation
because of Eq. (\ref{balance}) without requiring a change of the meridional 
flow.
A similar result can be obtained by lowering $\kappa_0$ and keeping 
$\delta_{\rm bc}$ constant. 

In panel b) and e) we show a solution with a factor of 2 larger width 
of the transition in the diffusivity profile $d_{\kappa\nu}$ (case 6). 
The most striking change is visible in the meridional flow pattern, where the 
return flow speed at the base of the convection zone is reduced by nearly a 
factor of 2. The amplitude of the differential rotation is only marginally
affected.

Panel c) and f) shows case 7 with the parameter $\alpha_{\kappa\nu}$ decreased
by a factor of 4 (value of the diffusivities at $r_{\rm tran}$).
The meridional flow pattern shows a reduction of the penetration into the 
subadiabatic layer because of the change in the Reynolds stress that drives 
this 
flow. The magnitude of the differential rotation at high latitudes decreases 
in response to the reduced spread of the entropy perturbation into the 
convection zone. Combining a lower value of $\alpha_{\kappa\nu}$ with a 
larger value of the subadiabaticity in the overshoot region as shown in 
panel (a) and c) would compensate for this effect and provide nearly
the same solution for the differential rotation (the penetration depth of 
the meridional flow would be still different).  

\begin{figure*}
  \resizebox{\hsize}{!}{\includegraphics{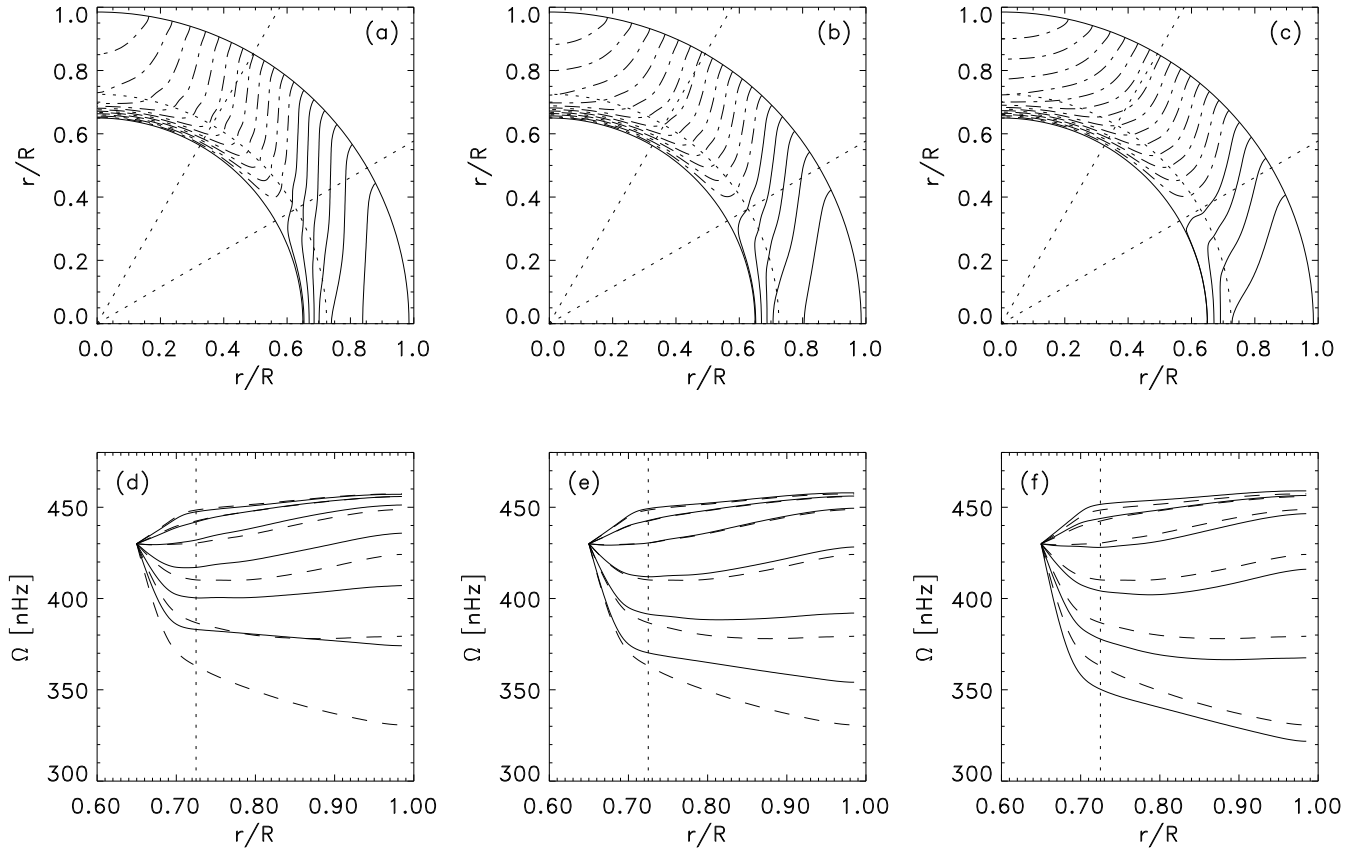}}
  \caption{Solutions with a nonadiabatic convection zone. (a, d)
    Case 8 with $r_{\rm sub}=0.8$, (b, e) case 9 with 
    $r_{\rm sub}=0.825$, and (c, f) case 10 with $r_{\rm sub}=0.85$.
    A superadiabatic convection zone can overcompensate for the effect of the 
    subadiabatic tachocline, leading again to cylindrical differential 
    rotation in the convection zone. However, if more than the lower third
    of the convection zone is weakly subadiabatic, this is sufficient
    to compensate for the upper superadiabatic part of the convection
    zone.}
  \label{f6}
\end{figure*}
   
So far we have discussed the influence of different profiles of viscosity
and thermal diffusivity, but the magnitudes set by $\nu_0$ and $\kappa_0$
have been left constant. We mentioned above that a change in $\kappa_0$
has the opposite effect as a change in $\delta_{\rm bc}$. Whereas an increase
in $\delta_{\rm bc}$ increases the entropy perturbation and through the
balance Eq. (\ref{balance}) the differential rotation, an increase in
$\kappa_0$ decreases both entropy perturbation and differential rotation.
An increase in $\nu_0$ increases the magnitude of the meridional flow
and, through the term $\sim \delta v_r$ in Eq. (\ref{entr}), also the 
entropy perturbation, which also results in an increase in differential 
rotation. However, if $\nu_0$ and $\kappa_0$ are changed together, the latter 
effect is compensated for by the larger thermal diffusivity and the 
differential 
rotation stays the same, whereas the meridional flow speed increases 
$\sim \nu_0$. These scalings change if $\nu_0$ becomes so large
that the viscous stress becomes a force comparable to the Coriolis force
in the meridional flow equation, or if $\kappa_0$ becomes so small that 
advection of entropy by the meridional flow dominates over the diffusion.

\section{Solutions with nonadiabatic convection zones} 
The models discussed so far assumed an adiabatic convection zone. This 
allowed us to separate the effects of a subadiabatic tachocline from
processes originating within the convection zone such as the rotational
anisotropy of convection, which are not considered in this model.

Since the meridional flow generates additional entropy variations
in a nonadiabatic convection zone, a consideration of a more realistic
stratification within the convection zone is crucial, especially since
a superadiabatic stratification generates entropy variations of
opposite sign. Since we are using an axisymmetric mean field approach,
incorporating a superadiabatic stratification has the potential to
introduce convective instability, which cannot be addressed appropriately
in this model. We therefore have to make sure that with the turbulent
viscosity and thermal diffusivity values we use, the Rayleigh number remains 
subcritical for convection.
For the superadiabaticity profile defined in Eq. (\ref{delta}) together with
a value of $\delta=3\times 10^{-5}$ at the top surface we have to increase
the thermal diffusivity in the upper part of the convection zone to values
of $5\times 10^9\,\mbox{m}^2\,\mbox{s}^{-1}$ to avoid convective instability.
We suppress here the convective instability only by increasing $\kappa_t$,
since a change of $\nu_t$ would also alter the turbulent angular momentum
transport and therefore change the differential rotation and meridional flow
pattern in a way that makes a comparison with the models discussed before
difficult. Since we only change $\kappa_t$ above $r_{\rm sub}$ the entropy 
perturbation that originates from the tachocline is only marginally
affected.

\begin{figure*}
  \resizebox{\hsize}{!}{\includegraphics{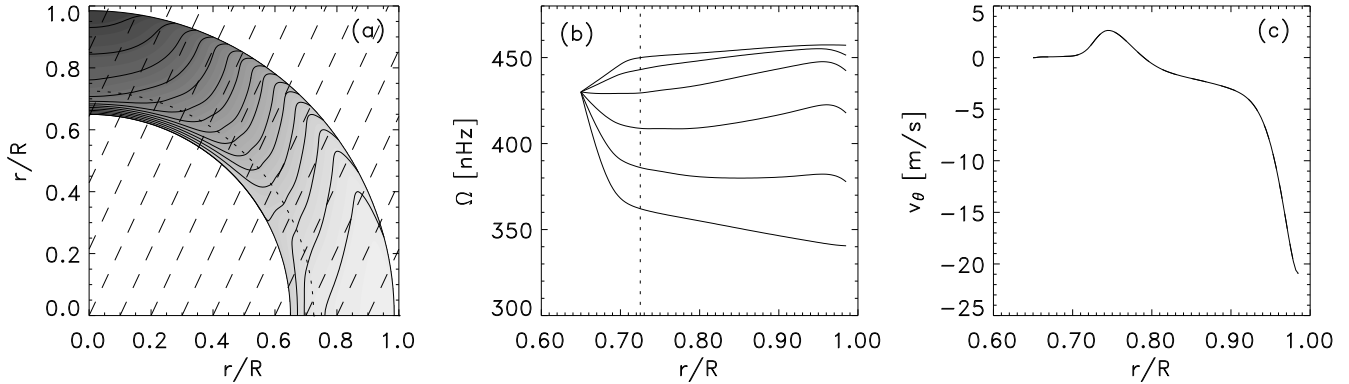}}
  \caption{Model for values of $r_{\rm sub}$ between those in cases 9 and 10,
    with a nonvanishing inward 
    angular momentum flux at the surface leading
    to a surface shear layer and an increased meridional return flow at
    the surface. In (a) we have overplotted dashed lines with a $25\degr$
    inclination with respect to the rotation axis, which is the observed
    inclination of the $\Omega$ contours between $15\degr$ and $55\degr$ 
    latitude.
  }
  \label{f7}
\end{figure*}

Nonlocal mixing-length models as discussed by \citet{Skaley:Stix:1991}
show typically below $r=0.75$ to $0.8\,R_{\odot}$ a weakly 
subadiabatic convection zone with values of $\delta\approx -5\times 10^{-7}$. 
At $r=0.95\,R_{\odot}$ the superadiabaticity reaches values around 
$\delta\approx 10^{-5}$ and increases then strongly toward the surface.
We cannot represent the large surface values for the reasons described above;
however, as we show later, they are also not that relevant for this problem. 

In Fig. \ref{f6} we shows models similar to case 1 but including
a nonadiabatic convection zone as defined by Eq. (\ref{delta}). We have
varied here the parameter $r_{\rm sub}$, which determines where the convection
zone turns weakly subadiabatic. In case 8 with 
$r_{\rm sub}=0.8\,R_{\odot}$
(panel a) and d)), the entropy perturbation arising from the superadiabatic
part of the convection zone is strong enough to overpower the effect of the
subadiabatic tachocline to a large extent. Below $45\degr$ latitude the
differential rotation is again close to the Taylor-Proudman state with
cylindrical contour lines; at higher latitudes the disk like rotation
contours are preserved, although the amplitude of the differential rotation
is significantly reduced. Using a value of $r_{\rm sub}=0.85\,R_{\odot}$
(case 10), the contribution of the subadiabatic part of the convection zone
dominates over the superadiabatic top part, and a solution close to that of
case 1
is retained (with slightly larger differential rotation). 
Comparing case 9 and case 10 shows that for the superadiabaticity
profile we have chosen, the transition takes place somewhere between
$r_{\rm sub}=0.825$ and $0.85\,R_{\odot}$.
 
We found in our model that the role of a nonadiabatic convection zone
depends on the extent of the weakly subadiabatic layer at the base. A 
transition in the solution takes place if roughly the lower $40\%-50\%$
of the convection zone
is subadiabatic, even though the subadiabaticity in the lower part is very 
weak. Before we discuss whether such a solution is feasible at all (see next 
paragraph), we investigate further how sensitively this result depends
on the assumption of the superadiabaticity profile within the convection zone
(e.g. we did not consider the strongly superadiabatic values close to the
top of the convection zone). 

Eq. (\ref{enthalpy}) expresses clearly that the large values
of the superadiabaticity near the surface do not matter that much, since 
the product $\rh_0\delta$ is relevant. The entropy perturbation 
at a given latitude in the convection zone results from an efficient diffusive 
exchange process of the quantity $\rh_0 T_0 s_1$ in radius, which means that
the lower part of the convection zone contributes significantly, even though
the value of $\vert\delta\vert$ is very small there. Consequently, if 
$\delta\sim\rh_0^{-1}$, which resembles the order of magnitude 
variation in mixing-length models, then the contribution of the lower 
half of the convection zone is roughly of magnitude equal to the upper half. 
Therefore, if the lower half of the convection zone is weakly subadiabatic, 
approximately $-10^{-6}$ to $ -10^{-7}$, this would dominate over the 
contribution of the superadiabatic upper part of the convection zone. 

Nonlocal mixing-length models do not provide such 
large subadiabatic fractions of the convection zone. \citet{Spruit:1997}
suggested a highly nonlocal convection model, in which the convection is 
driven 
in a narrow superadiabatic surface layer and downflows penetrate with a small
amount of mixing all the way to the base of the convection zone. In this case
the major fraction of the convection zone would be subadiabatic (because of
radiative heating of the broad upflow in the lower convection zone), but it 
is unclear to what extent this model applies to solar convection. 
  
In Fig. \ref{f7} we show a solution with a value of $r_{\rm sub}$ between
the values used for case 9 and 10 ($r_{\rm sub}=0.8375\,R_{\odot}$) 
and a different boundary condition for
the angular velocity at the surface. Instead of setting the turbulent
angular momentum transport to zero as expressed in Eq. (\ref{amflux}),
we use a radially inward transport by setting $\lambda=-\theta$ in a thin
surface layer. Because of the stress-free boundary condition for $\Omega$ this
requires a negative radial gradient of $\Omega$ near the surface. This
parameterization reflects the idea by \citet{Gilman:Foukal:1979} that the 
rotational influence on supergranulation leads to an outwardly decreasing
$\Omega$.

Fig. \ref{f7} panel a) shows the differential rotation contours for this
case. Over plotted are lines indicating a $25\degr$ inclination angle with 
respect
to the axis of rotation as found by helioseismology 
\citep{Schou:etal:1998,Schou:etal:2002}. This feature is reproduced very well
from about $15\degr$ to about $60\degr$ latitude. \citet{Gilman:Howe:2003} 
presented 
an explanation for this phenomenon based on the influence of a one-cell
meridional flow on the differential rotation, assuming that the differential
rotation would be constant in radius in the absence of the meridional flow.
In our model it is difficult to identify the solution that would correspond
to a solution in the absence of the meridional flow since the meridional
flow is an integral part of the solution. Neglecting any effect of the
meridional flow would provide a solution in which the Reynolds stress
is balanced by diffusion, which in this particular case would yield
contour lines with a $75\degr$ angle to the axis of rotation. In that sense
the inclination is a result of the meridional flow but also including 
in a more complicated way the influence on the entropy profile
within the convection zone. 
 
In panel c) we show the meridional
flow speed at $45\degr$ latitude. A comparison with Fig. \ref{f4} shows an 
increase in the surface flow speed, which is a direct result of the changed
boundary condition for the angular velocity.

We present this solution here in order to show that within the framework
of this model it is possible to obtain solutions that are very close to the 
observed pattern by making assumptions about the superadiabaticity that
go beyond the predictions of mixing-length theory (subadiabatic part of 
convection zone extends up to $0.8375\,R_{\odot}$), but which are physically 
feasible if the degree of nonlocality of the convection is large enough.  

\section{Summary}
The main results of our model are as follows:
\begin{enumerate}
  \item The profile of the differential rotation is determined mainly
    by the profile of the entropy perturbation originating in the subadiabatic
    tachocline and spreading into the convection zone because of thermal 
    conductivity.
  \item The profile of the differential rotation is rather insensitive to
    the parameterization of the Reynolds stress as long as there is a 
    sufficiently large equatorward angular momentum flux; however, the
    magnitude of differential rotation changes with different assumptions.
  \item The parameterization of the Reynolds stress strongly influences the
    meridional flow (compare cases 1, 2, 6, and 7).
  \item If the lower half of the convection zone is weakly subadiabatic, the 
    solar differential rotation can be explained through entropy 
    perturbations arising from the nonadiabatic stratification. If the
    subadiabatic region has a smaller extent, additional effects such as
    anisotropic heat transport are required.
  \item For angular momentum transport almost aligned with the axis of rotation
    and angular momentum transport in latitude only, we find a dominant 
    counterclockwise meridional flow cell (equatorward at the base of and
    poleward at the upper layers of
    of convection zone) as a robust result. Our model shows also the tendency
    of a weaker reverse cell at higher latitudes, which has been observed
    by \citet{Haber:etal:2002}. A dominant counterclockwise meridional flow 
    cell is favorable for flux-transport dynamo models, in which 
    the equatorward
    meridional flow at the base of the convection zone ensures the equatorward
    propagation of magnetic  activity through the solar cycle.
  \item The meridional flow shows very little penetration beneath the base of 
    the convection zone. In addition to the constraints imposed by the 
    subadiabatic stratification in the radiative interior, angular momentum
    conservation does not allow for a significant meridional flow there, since 
    the associated angular momentum transport would lead to significant changes
    of differential rotation unless opposed by a strong Reynolds stress 
    (which is highly unlikely in the radiative interior).
\end{enumerate}

\section{Implication for solar differential rotation}
We presented in this paper a simplified model for the solar differential
rotation and meridional flow. This model parametrizes important convective 
scale processes such as the turbulent angular momentum transport and 
turbulent diffusivities and does exclude processes such as  rotational
anisotropy of the convective energy flux as discussed in detail by
\citet{Kitchatinov:Ruediger:1995} and \citet{Kueker:Stix:2001}. This model
is therefore not intended to be a complete differential rotation model
for the solar convection zone, but rather a model to evaluate the
importance of effects resulting from a nonadiabatic stratification in
tachocline and convection zone. 

Even though we did not use here exactly the formulation of the $\Lambda$-effect
as adopted by \citet{Kitchatinov:Ruediger:1995}, \citet{Ruediger:etal:1998},
and \citet{Kueker:Stix:2001} for the solar case, our results are in general
agreement with their earlier work. For a magnitude of the $\Lambda$-effect
(the parameter $\Lambda_0$ in our model) of order unity,
we get an amplitude of the differential rotation comparable to solar values.
Differences occur in the profile of the differential rotation (inclination
of isolines with respect to the axis of rotation) because of the different 
physics considered in the entropy equation of our model (which is the main 
focus of this work). Larger, but explainable differences, exist in the 
meridional flow patterns obtained. Although the investigation of 
\citet{Ruediger:etal:1998} shows 
a counterclockwise flow similar to our results in a model including a 
significant amount of viscosity and no latitudinal entropy variation, models 
including a latitudinal entropy gradient and a $\Lambda$-effect formulation 
taking into account the variation of the Coriolis number within the convection 
zone typically yield a clockwise flow cell close to the surface 
\citep{Kitchatinov:Ruediger:1995,Kueker:Stix:2001}. As explained by the authors
this is because of a radially outward transport of angular momentum in the 
upper layers, which is not parameterized in our model. For most of our 
cases we assume
a vanishing angular momentum flux at the surface; in the case shown in Fig. 
\ref{f7} we use a radially inward transport of angular momentum 
(as suggested by the 
work of \citet{Gilman:Foukal:1979} and also by observations, which clearly 
show a decrease of rotation rate with radius in the outer layers), which has 
the opposite effect of enhancing the poleward meridional surface flow.
The investigation of \citet{Kueker:Stix:2001} also indicates that the 
meridional flow
is much more sensitive to details of the model used than the differential
rotation (see Fig. 2 therein, which addresses the influence of the 
mixing-length parameter), in agreement with the findings in this 
paper.    
    
The models with adiabatic convection zones, which are shown in Fig. \ref{f2}
to \ref{f5}, allow evaluating of the importance of the tachocline effects
for models such as that of \citet{Kueker:Stix:2001} that only consider the 
convection
zone. To summarize our results, the entropy perturbation generated within
a subadiabatic tachocline is strong enough to avoid the Taylor-Proudman
state above $30\degr$ latitude and is therefore as important as other
effects such as rotational anisotropy of the convective energy transport.
However, \citet{Kitchatinov:Ruediger:1995} found in their investigation
that the effect of a subadiabatic tachocline is rather small compared with the
effect of anisotropic energy transport. This apparent contradiction
could be caused either by the different steepness of the shear
profile of the differential rotation in the tachocline in the model of
\citet{Kitchatinov:Ruediger:1995} (see, e.g., Fig. 1 and 2 therein), since the 
magnitude of the entropy perturbation depends strongly on the value of 
$\partial \Omega/\partial z$, or by an insufficient overlap between thermal
conductivity profile and subadiabatic overshoot region 
(\citet{Kitchatinov:Ruediger:1995} computed the turbulent diffusivity through 
a local mixing-length relation). 

Since the value of $\partial \Omega/\partial z$
within the subadiabatic tachocline is significantly larger than the
value of $\partial \Omega/\partial z$ in the convection zone, the magnitude
of the expected entropy perturbation in the tachocline also exceeds the
magnitude required to balance a solar like differential rotation within
the convection zone. If therefore only a fraction of the 
entropy perturbation generated within the subadiabatic tachocline spreads
into the convection zone, this provides a significant contribution.
This spread depends mainly on the overlap between the subadiabatic tachocline
and the region that is mixed by convection. In our model we typically use a
convective diffusivity of $5\times 10^7\mbox{m}^2\,\mbox{s}^{-1}$ at
$r=r_{\rm tran}=0.725\,R_{\odot}$ and lower values below, which introduced 
enough coupling between the subadiabatic region and the convection zone.
Since the observed tachocline spreads about one-third of the way into the 
region above $r=0.713\,R_{\odot}$, which is very close to adiabatic according
to helioseismology \citep{Charbonneau:etal:1999}, it can be expected that 
there is a sufficient coupling between these two regions in the case of the 
Sun.

The models with a nonadiabatic convection zone (Figs. \ref{f6}, \ref{f7})
allow us to estimate under which conditions the entropy perturbation 
resulting from the nonadiabatic stratification is sufficient to explain
the observed solar differential rotation without any additional effects
such as anisotropic heat transport. The main problem is that a
superadiabatic convection zone overcompensates the effect of the 
subadiabatic tachocline to a large degree unless the lower half of the 
convection zone is weakly subadiabatic. Even though the absolute values of the
superadiabaticity in the convection zone are much lower than those in the
tachocline, their contribution can be very large because of the much larger
radial meridional flow velocity within the convection zone. We found that the 
effect of the superadiabatic part of the convection zone can be compensated
by a weakly subadiabatic lower half of the convection zone, since the thermal
inertia of the entropy perturbation is $\sim \rh_0 T_0 s_1$ and therefore even
a small entropy perturbation at the base of the convection zone can contribute
more than a large entropy perturbation in the upper part of the
convection zone. Within the frame work of this model we cannot address the
question of whether such a solution is feasible, since this would require 
more sophisticated nonlocal convection theory. Nonlocal mixing-length
models typically predict a subadiabatic stratification below 
$r=0.8\,R_{\odot}$;
however, for a larger degree of nonlocality (stable downflows with only
little entrainment and detrainment) a larger extent of this region would be
possible. If additional effects such as anisotropic heat transport and 
nonlinear
feedback of the stratification on the heat conductivity were included, it could
be possible that a solar-like solution as shown in Fig. \ref{f7} is possible
with a smaller extent of the subadiabatic part of the convection zone.

A further test of the process proposed in this paper would be the inclusion
of a subadiabatic tachocline in the full spherical shell convection 
simulations. Besides adding a subadiabatic overshoot region, this requires 
the inclusion of the tachocline shear layer either by resolving the
relevant physical processes or by forcing the shear (as done in this model)
through a lower boundary condition, since the entropy perturbation arises
as a consequence of a subadiabatic shear layer. Since the entropy perturbation
is also affected significantly by the meridional flow within a nonadiabatic
convection zone, it is also crucial to get the meridional flow
pattern right. Work in this direction is currently in progress.

Since the meridional flow is of high interest for flux transport dynamos,
here we discuss in more detail the findings of our model concerning the
penetration of the return flow below the base of the convection zone.
As mentioned above, our model provides a 
counterclockwise flow as a robust result, provided that there is an inwardly 
directed turbulent angular momentum flux. The return flow at the base of
the convection zone is located in the region with the largest turbulent
viscosity gradient, since there the divergence of the turbulent angular
momentum flux ($\Lambda$-effect) would lead to an increase of rotation rate
unless opposed by an equatorward transport of angular momentum by the 
meridional flow (which tends to slow down the rotation rate). In that way the 
meridional flow is always tied to the presence of turbulent Reynolds stress. 
For the parameterization of the turbulent diffusivities we use in most
of our models, the meridional flow speed at $r=0.71\,R_{\odot}$ typically 
drops to less than $10\%$ of the maximum return flow speed, which is reached 
at around $r=0.745\,R_{\odot}$. Below $r=0.71\,R_{\odot}$ velocities on the 
order of a few $\mbox{cm}\,\mbox{s}^{-1}$ persist, however our model tends 
to overestimate
their amplitude since we use overshoot values for $\delta$. Using radiative 
core values would decrease this amplitude further by several orders of 
magnitude. Therefore, a penetration of the meridional flow below the base of 
the convection zone as used in a few flux transport dynamo models 
seems very unreasonable, since it implies that there are very strong 
Reynolds stresses in the strongly subadiabatic radiative core. Since the 
constraint
set by the angular momentum conservation is additional to the constraint set
by the subadiabaticity of the stratification, which was pointed out
by \citet{Gilman:Miesch:2004}, the limitation on penetration
of meridional flow beneath the base of the convection zone is even more 
stringent than found by \citet{Gilman:Miesch:2004}.  
     
\acknowledgements
The author wants to thank P.~A. Gilman and M.~S. Miesch for stimulating 
discussions about the solar differential rotation problem.

\bibliographystyle{../../../natbib/apj}
\bibliography{../../../natbib/apj-jour,../../../natbib/papref}

\end{document}